\documentclass[reprint,ltxfront,twocolumn,secnumarabic,amssymb, nobibnotes,aps,floatfix,10pt,superscriptaddress]{revtex4-2}

\usepackage{float}
\usepackage{graphicx}
\usepackage{amsmath}
\usepackage{amsfonts}
\usepackage{stmaryrd}
\usepackage{bbm}
\usepackage{mathrsfs}
\usepackage{tipa}
\usepackage{amssymb}
\usepackage{txfonts}
\usepackage{dcolumn}
\usepackage{epstopdf}
\usepackage{array}
\usepackage{longtable}
\usepackage{subfigure}
\usepackage{multirow}
\usepackage[colorlinks=true,linkcolor=black,citecolor=blue,urlcolor=blue,]{hyperref}
\usepackage[doipre={doi:~}]{uri}
\usepackage{CJK}
\usepackage{indentfirst}
\usepackage{cases}
\usepackage{hyperref}
\hypersetup{hypertex=true,
	colorlinks=true,
	linkcolor=blue,
	anchorcolor=blue,
	citecolor=blue,
	urlcolor=blue,
	menucolor=blue}
	
\newcommand*{\cm}{cm$^{-1}$}
\newcommand*{\R}{$\Delta R/R$\ }
\newcommand*{\Ag}{$A_{1g}$\ }
\newcommand*{\uJcm}{$\muup$J/cm$^2$\ }
\newcommand*{\eph}{e-ph}
\newcommand*{\eh}{e-h }
\newcommand*{\F}{$F$}
\newcommand*{\T}{$T$}

\begin{document}

\title{Nonequilibrium carrier and phonon dynamics in the ferrimagnetic semiconductor Mn$_3$Si$_2$Te$_6$}

\author{Y. Yang}
\affiliation{Key Laboratory of Education Ministry for Laser and Infrared System Integration Technology, Shandong University, Qingdao 266237, China}
\affiliation{Beijing National Laboratory for Condensed Matter Physics, Institute of Physics, Chinese Academy of Sciences, Beijing 100190, China}
	
\author{X. T. Chen}
\author{Z. L. Li}
\affiliation{Beijing National Laboratory for Condensed Matter
	Physics, Institute of Physics, Chinese Academy of Sciences, Beijing 100190, China}

\author{J. B. Pan}
\affiliation{Beijing National Laboratory for Condensed Matter
	Physics, Institute of Physics, Chinese Academy of Sciences, Beijing 100190, China}
	
\author{F. Jing}
\affiliation{Beijing National Laboratory for Condensed Matter
		Physics, Institute of Physics, Chinese Academy of Sciences, Beijing 100190, China}

\author{S. S. Zhang}
\email{sasazhang@sdu.edu.cn}
\affiliation{Key Laboratory of Education Ministry for Laser and Infrared System Integration Technology, Shandong University, Qingdao 266237, China}
\affiliation{School of Information Science and Engineering, Shandong University, Qingdao 266237, China}	

\author{X. B. Wang}
\email{xinbowang@iphy.ac.cn}
\affiliation{Beijing National Laboratory for Condensed Matter
	Physics, Institute of Physics, Chinese Academy of Sciences, Beijing 100190, China}

\author{J. L. Luo}
\affiliation{Beijing National Laboratory for Condensed Matter
	Physics, Institute of Physics, Chinese Academy of Sciences, Beijing 100190, China}
	
\date{\today}

\begin{abstract}
We investigate the ultrafast carrier and phonon dynamics in the ferrimagnetic semiconductor Mn$_3$Si$_2$Te$_6$ using time-resolved optical pump-probe spectroscopy. Our results reveal that the electron-phonon thermalization process with a subpicosecond timescale is prolonged by the hot-phonon bottleneck effect. We identify the subsequent relaxation processes associated with two non-radiative recombination mechanisms, i.e., phonon-assisted electron-hole recombination and defect-related Shockley-Read-Hall recombination. Temperature-dependent measurements indicate that all three relaxation components show large variation around 175 and 78 K, which is related to the initiation of spin fluctuation and ferrimagnetic order in Mn$_3$Si$_2$Te$_6$. In addition, two pronounced coherent optical phonons are observed, in which the phonon with a frequency of 3.7 THz is attributed to the \Ag mode of Te precipitates. Applying the strain pulse propagation model to the  coherent acoustic phonons yields a penetration depth of 506 nm and  a sound speed of 2.42 km/s in Mn$_3$Si$_2$Te$_6$. Our results develop understanding of the nonequilibrium properties of the ferrimagnetic semiconductor Mn$_3$Si$_2$Te$_6$, and also shed light on its potential applications in optoelectronic and spintronic devices.
\end{abstract}

\maketitle

\section{Introduction}
Topological materials host symmetry-protected band crossings in momentum space, leading to intriguing electronic and transport properties   \cite{hasan_weyl_2021}. Among various topological materials, topological nodal line materials are characterized by crossed bands forming lines or rings   \cite{PhysRevX.8.031044}. The layered compound Mn$_3$Si$_2$Te$_6$ (MST) is a newly discovered topological nodal line material   \cite{seo2021colossal}, and thus there has been increased interest in studying this compound. MST is a ferrimagnet below the Curie temperature of $T_c$ $\sim$78 K with a tilted spin configuration toward the c axis   \cite{olmos2021critical,liu2018critical,may2017magnetic,Spin160K}.  The first-principles calculations revealed an indirect bandgap and the topological nodal line structure of the valence Te band in MST  \cite{seo2021colossal,zhang2023electronic}. The topological nodal line degeneracy can be lifted depending on spin orientation. When the magnetic field is applied along the hard axis, the compound undergoes an insulator-metal transition, leading to the colossal magneto resistance (CMR) effect \cite{ni2021colossal,seo2021colossal}.  The anomalous Nernst effect and possible topological Nernst effect are found in the low-field region by electrical and thermoelectric transport measurements \cite{ANE2023}. Recent work also demonstrated current-control of chiral-orbital-current-enabled CMR in MST which offers a new paradigm for quantum technologies \cite{Zhang2022}. Therefore, the magnetic semiconductor MST is a promising option for next-generation optoelectronic and spintronic devices. Clarifying the mechanisms of nonequilibrium quasiparticles relaxation in MST is crucial not only for studying fundamental physics in topological nodal line materials \cite{Kirby2021ZrSiTe,yu2020ultrafast,Biswas2022ZrSiS} but also for potential future applications \cite{2021Ultrasensitive}. 

Time-resolved ultrafast pump probe spectroscopy is a powerful tool to address nonequilibrium dynamics by disentangling distinct quasiparticle decay processes after photo-excitation \cite{seifert2022ultrafast,dean2016ultrafast,hu2022optical}. It has been successfully employed to trace the decay processes of nonequilibrium quasiparticles in topological materials \cite{Snapshots2013,Selective2011}. Relaxation mechanisms of photo-excited quasiparticles usually involve a complex interplay of carrier-carrier, carrier-phonon, and phonon-phonon interactions that are dominated by the band structure \cite{Gierz2013}.

In this work, we carry out a detailed ultrafast transient reflectivity study of MST single crystals by varying temperature and pump fluence. Our results reveal that the fast relaxation process with a subpicosecond timescale originating from the electron-phonon (e-ph) thermalization process is prolonged by the hot phonon bottleneck effect. Two slow carrier relaxation processes with different timescales are attributed to the phonon-assisted electron–hole (e-h) recombination and trap-assisted Shockley-Read-Hall (SRH) recombination. The temperature dependent experiments indicates that the three relaxation components show variations around 175 and 78 K, which is associated with the evolution of magnetic interaction. Furthermore, two distinct optical phonons and one pronounced acoustic phonon are observed. Combined with the Raman scattering measurement, the coherent optical phonon with a frequency of 3.7 THz is attributed to the \Ag mode of Te precipitates whose temperature dependence of the frequency and damping rate can be well described by an anharmonic model. The strain pulse propagation model is applied to explain coherent acoustic phonons, yielding a penetration depth of 506 nm and a sound speed of 2.42 km/s. These findings not only provide critical insights into the nonequilibrium properties of Mn$_3$Si$_2$Te$_6$ but also shed light on the potential applications of this magnetic semiconductors in optoelectronic and spintronic devices.

\section{Experimental Methods}

MST single crystals were grown by the chemical vapor transport (CVT) method using a presynthesized compound as the precursor and iodine (5 mg/mL) as a transport agent. The temperature gradient was 800 $^{\circ}$C to 750 $^{\circ}$C and the CVT lasted for 2 weeks. Single crystals with a size of $\sim$5 mm with a regular shape and shiny surface were finally obtained.  In the transient reflectivity measurements, an optical parametric amplifier (Orpheus F, Light Conversion) was pumped by the output of a high-repetition-rate Yb:KGW amplifier (Pharos, Light Conversion) to generate ultrafast laser pulses with a center wavelength of 800 nm, a pulse duration of 40 fs, and a repetition rate of  50 kHz. The cross-polarized pump and probe beams were focused on the $ab$ face of the MST crystal with spot diameters of 100 and 50 ${\muup}$m, respectively. The probe beam was horizontally polarized with an incident angle $\sim \pi/12$. The sample was mounted in a continuous-flow liquid helium cryostat with temperature varying from 10 to 300 K.

\section{Experimental Results}

\begin{figure}[b]
	\includegraphics[width=3.2in]{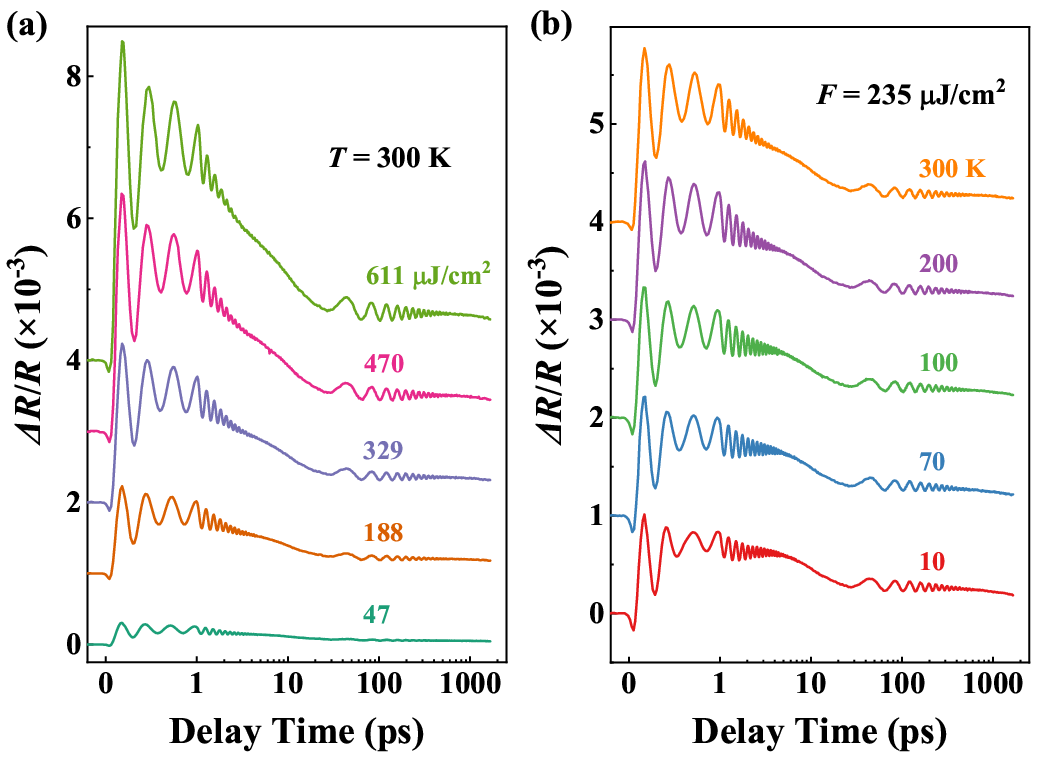}
	\subfigure{\label{1a}}
	\subfigure{\label{1b}}
	\caption{(a) \R in MST measured with different pump fluences at \T = 300 K. (b) \R measured at selected temperatures between 10 and 300 K with pump fluence kept at \F = 235 \uJcm. The curves are shifted vertically for clarity.}
	\label{1}
\end{figure}

\begin{figure}[tb]
	\includegraphics[width=3.2in]{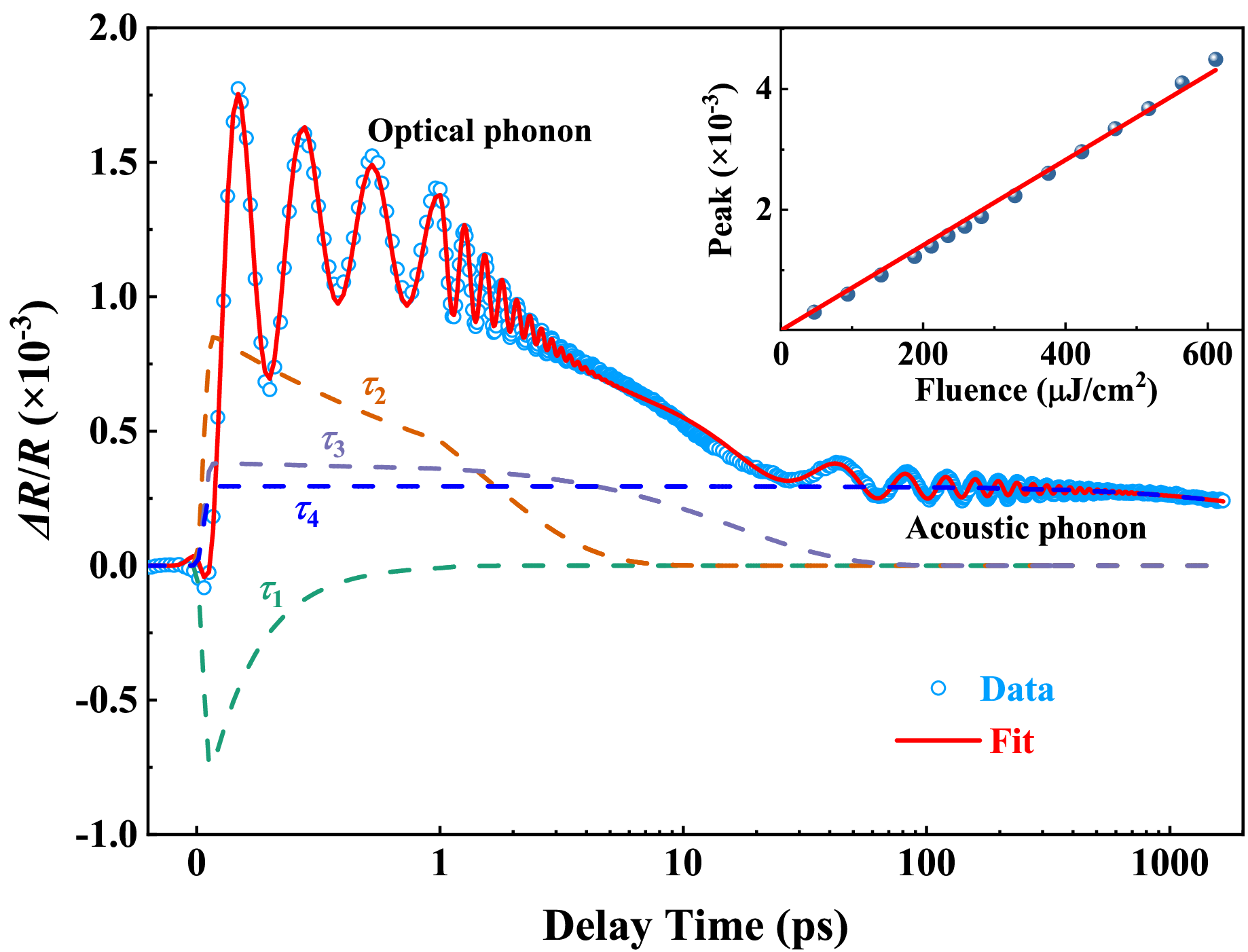}
	\caption{ Typical fitting of \R at \T = 300 K and \F = 235 \uJcm using Eq. \eqref{1}. The  dashed lines represent four carrier relaxation processes. The inset shows the fluence-dependent peak values, where the solid line is a linear fit.}
	\label{2}
\end{figure}

The transient differential reflectivity is defined as the pump-induced relative change of the probe reflectivity, i.e., \R = $[R(t)-R]/R$. Figures \ref{1a} and \ref{1b} show the pump fluence \F and temperature \T dependence of the \R signals, respectively. All the curves are plotted on a logarithmic time-scale and shifted vertically for clarity.  The \R signals exhibit an initial small dip, followed by an abrupt rise to a peak, and then undergo a multi-exponential decay process accompanied by distinct periodic oscillations. The peak amplitude increases linearly with fluence, as indicated in the inset of Fig. \ref{2}, indicating that the optical transitions were not saturated in our experiments. During the temperature-dependent measurements, the pump fluence was kept at 235 $\muup$J/cm$^2$. Note that a discernible phase shift occurs around 0.6 ps on the 10 K curve in Fig. \ref{1b}, indicating the potential existence of two high-frequency oscillatory components in the first 10 ps. Additionally, another slow oscillatory component with a period of 38 ps emerges at 20 ps and extends to around 500 ps. To gain quantitative insights into the rich dynamics, the \R signal is fitted using the following formula \cite{TCL2002,BSTS2014}:
\begin{equation}\label{E1}
	\begin{aligned}
	%\begin{split}
\frac{{\Delta R}}{R} = [\sum_{i = 1}^4 {{A_i}} {e^{\frac{{ - t}}{{{\tau _{i}}}}}} + \sum_{j = 1}^3 {{B_j}} {e^{{ -{{\gamma}_{j}} t}}\sin (2\pi {f_{j}}t + {\varphi _j})}] \otimes G({t}),
	\end{aligned}
	%\end{split}
\end{equation}
where $A_i$ and $\tau_{i}$ are the amplitude and relaxation time of the $i$th decay component, respectively; $B_j$, $\gamma_{j}$, $f_{j}$ and $\phi_j$ are the amplitude, damping rate, frequency and  phase of the $j$th oscillatory component, respectively. $G(t)$ is a Gaussian function representing the pump-probe cross correlation.  The transient reflectivity data can be well fitted with Eq. \eqref{E1} at all measured temperature and fluence.

\begin{table}[bp]
	\caption{\label{tab:table1}
	Relaxation times of the carrier dynamics and the extracted phonon frequencies at 300K.}
	\begin{ruledtabular}
	\begin{tabular}{cccc}
	\multicolumn{2}{c}{Relaxation time (ps)} & \multicolumn{2}{c}{Phonon frequency (THz)} \\
	\hline
	$\tau_{1}$ & 0.21 ±0.01 & $f_{1}$ & 3.7 ±0.01\\
	$\tau_{2}$ & 1.52 ±0.03 & $f_{2}$ & 4.3 ±0.04\\
	$\tau_{3}$ & 16.75 ±0.33 & $f_{3}$ & 0.026±3$\times$$10^{-5}$\\
	$\tau_{4}$ & 7840 ±319.34 &   &  \\
	\end{tabular}
	\end{ruledtabular}
\end{table}

\begin{figure*}[tbp]
	\centering
	\includegraphics[width=6.4 in]{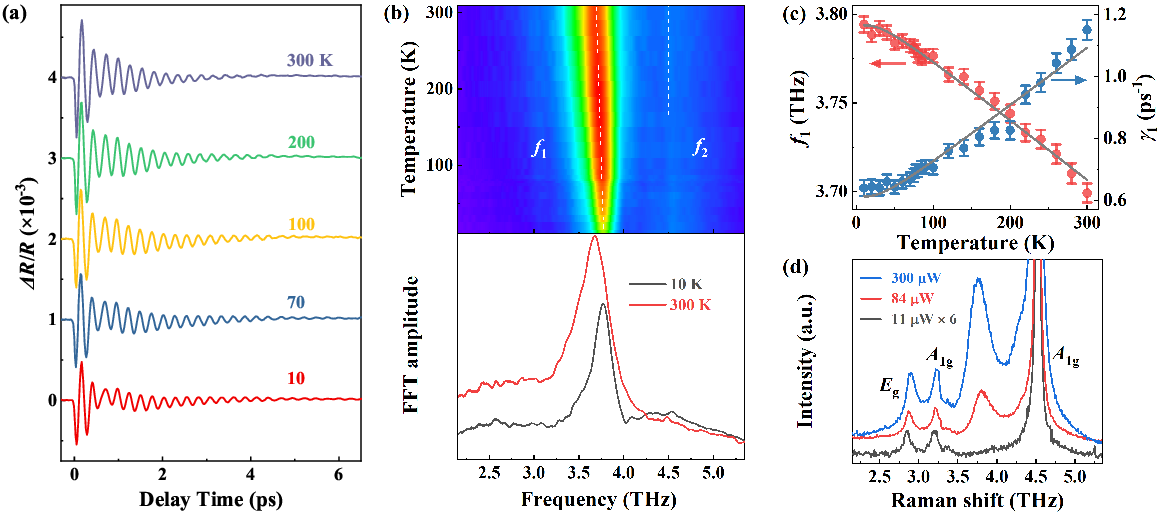}
	\subfigure{\label{5a}}
	\subfigure{\label{5b}}
	\subfigure{\label{5c}}
	\subfigure{\label{5d}}
	\caption{(a) Coherent phonon oscillations after subtracting the electronic background for selected temperatures.
		(b) Fast Fourier transformation of the subtracted data as a function of temperature in the frequency domain.
		(c) Temperature-dependent phonon frequency $f_1$ and damping rate $\gamma_{1}$. The solid lines represent the fitting curves using Eqs. \eqref{E4} and \eqref{E5}, respectively. (d) The Raman spectra measured at different laser power levels. The spectra are shifted vertically for the sake of clarity. The spectra under the 11 $\muup$W level are multiplied by a factor of 6.}
	\label{5}
\end{figure*}

As a representative example at 300 K, the fitting curve in Fig. \ref{2} can reproduce the experimental data quite well. The extracted relaxation times of the exponential decay components and oscillating frequencies are presented in Table \ref{tab:table1}.
As will be discussed in more details below, the four decomposed decaying components as indicated by dashed lines in Fig. \ref{2} are related to the recovery dynamics of photo-excited carriers. Among them, the slowest decay component with a timescale of several nanoseconds is related to heat diffusion out of the excitation volume, which will not be discussed in this paper. Generally, the oscillatory components with terahertz frequency in the transient reflectivity arise from the coherent optical phonons. To further confirm the reliability of our fitting, the oscillations from \R signals were extracted by subtracting the exponential decay components and then performing a fast Fourier transform (FFT). It can intuitively be seen in Fig. \ref{5b} below that a sharp peak at 3.7 THz and a broad hump around 4.5 THz exist. These two peaks of the coherent optical phonons are consistent with the fitted frequencies, supporting the robustness of our fitting. In addition, the low-frequency oscillatory component with a frequency of 0.026 THz is attributed to a coherent acoustic phonon. In the following, we will discuss the carrier and phonon relaxation dynamics in succession.

\subsection{Ultrafast carrier dynamics}

\begin{figure}[tbp]
	\includegraphics[width=3.2 in]{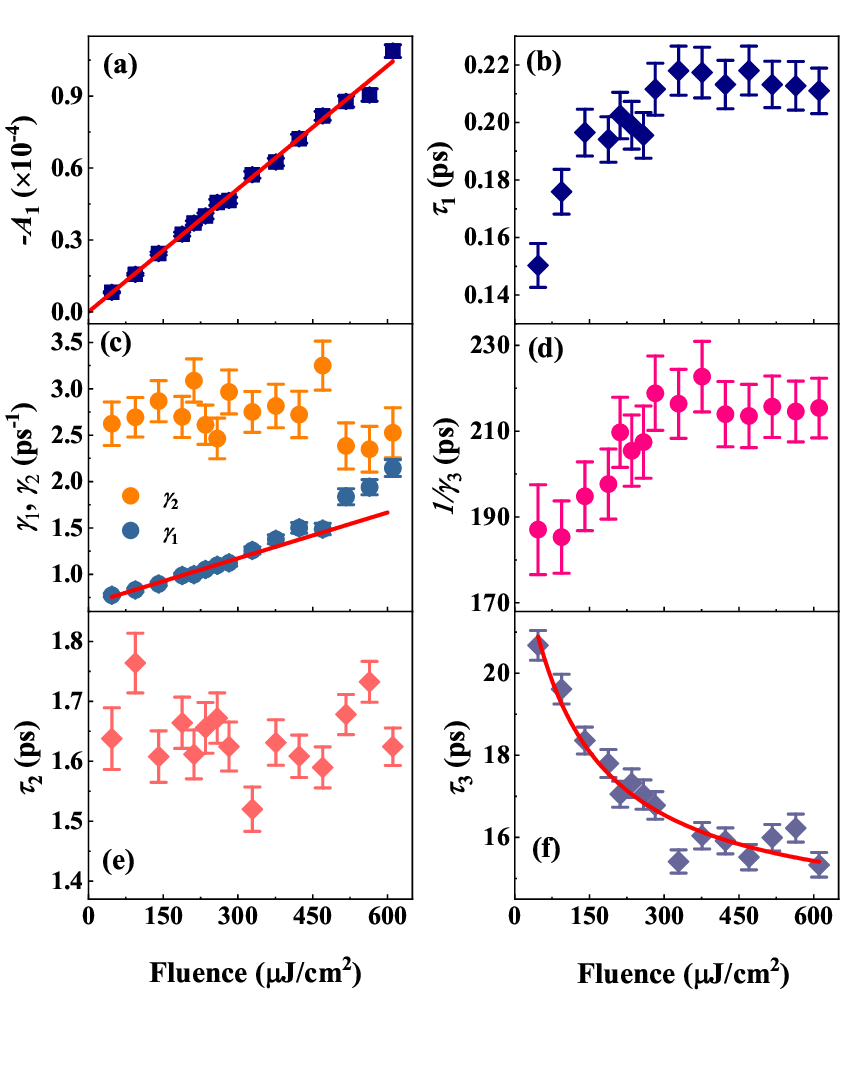}
	\subfigure{\label{3a}}
	\subfigure{\label{3b}}
	\subfigure{\label{3c}}
	\subfigure{\label{3d}}
	\subfigure{\label{3e}}
	\subfigure{\label{3f}}
	\caption{Fluence dependence of (a) the amplitude -$A_1$, (b) relaxation time $\tau_{1}$, (c) damping rate of the coherent optical phonons, (d) lifetime (reciprocal damping rate) of the acoustic phonon at 300 K, (e) relaxation time $\tau_{2}$, and (f) relaxation time $\tau_{3}$. The solid curves in (a) and (c) are linear fits. The solid line in (f) is the fitting result using Eq. \eqref{E3}.}
	\label{3}
\end{figure}

First, we focus on the fast decay process characterized by a sub-picosecond timescale and negative amplitude. Figures \ref{3a} and \ref{3b} show the amplitude -$A_1$ and relaxation time $\tau_{1}$ as a function of pump fluence, respectively. The absolute amplitude increases linearly with the pump fluence in the whole studied range, indicating that there is no saturation of photo-excited carrier concentration. $\tau_{1}$ increases monotonically with pump fluence up to a distinct threshold $F^*\sim$330 $\muup$J/cm$^2$, after which it reaches a constant at higher excitation densities. In general, after electron thermalization, the energetic carriers cools by dissipating excess energy to the lattice via \eph interaction. A similar relaxation process with negative amplitude has been observed in other topological materials, such as NiTe$_2$ \cite{cheng2022ultrafast} and TaAs \cite{wu2020quasiparticle}. Therefore, we ascribe the $\tau_{1}$ process to the relaxation of the carrier from a high energy level to the bottom of the band via \eph coupling. It is known that the fast decay component usually reflects the \eph coupling strength. The fast relaxation time indicates the relatively strong \eph interaction in MST, as evidenced by the pronounced oscillations appearring when the \R signal reaches its maximum value.

The relaxation of hot carriers primarily depends on the emission of optical phonons. The generated optical phonon can decay into two acoustic phonons via anharmonic phonon-phonon interactions \cite{Hot2020ZrTe5,Anh2008}. When the relaxation time of the optical phonons is comparable to or longer than the hot carriers cooling time, an appreciable nonequilibrium phonon (hot phonon) population accumulates \cite{SENNA1987,Huang2010}. Such hot phonons may reheat the cooled electrons and hence retard the hot carriers relaxation processes, which manifests as the hot-phonon bottleneck effect that is commonly observed in highly excited polar semiconductors \cite{Shah1999,2017Hot}.  Under high excitation, $\tau_{1}$ increases with pump fluence due to the accumulation of hot phonons. According to Eq. (\ref{E4}), the damping rate of the optical phonon increases with its occupation number. As shown in Fig. \ref{3c}, $\gamma_{1}$ increases with pump fluence below $F^*$. On the contrary, $\gamma_{2}$ is approximately fluence independent, which is probably due to the fact that the $f_2$ optical phonon mode interacts with the hot carriers in the early stage of thermalization, in which a large non-equilibrium phonon population has not yet been achieved. Furthermore, the lifetime (reciprocal damping rate) of the acoustic phonon increases with fluence as illustrated in Fig. \ref{3d}. Generally, a coherent acoustic phonon is induced by the electronic and/or thermal stress at the photo-excited sample surface leading to strain pulse propagation into the sample \cite{thomsen1986surface}. In MST, the hot phonons can be reabsorbed by the photo-carriers leading to a slower attenuation rate of the stress pules and thus an increase of the acoustic phonon lifetime. At very high photoexcitation density, the hot carriers and the optical phonons reach a quasi-equilibrium state, in which the relaxation time of the hot carriers is limited by that of the optical phonons \cite{Huang2010}. As shown in Fig. \ref{3b}, $\tau_1$ reaches its constant value of  $\sim$0.21 ps, which is roughly half of the two optical phonon lifetimes above $F^*$. In addition, the lifetime of the acoustic phonon also exhibits saturation behavior, reminiscent of $\tau_{1}$. These results suggest the presence of the fluence-dependent hot-phonon bottleneck effect in MST.

\begin{figure}[tb]
	\centering
	\includegraphics[width=3.2in]{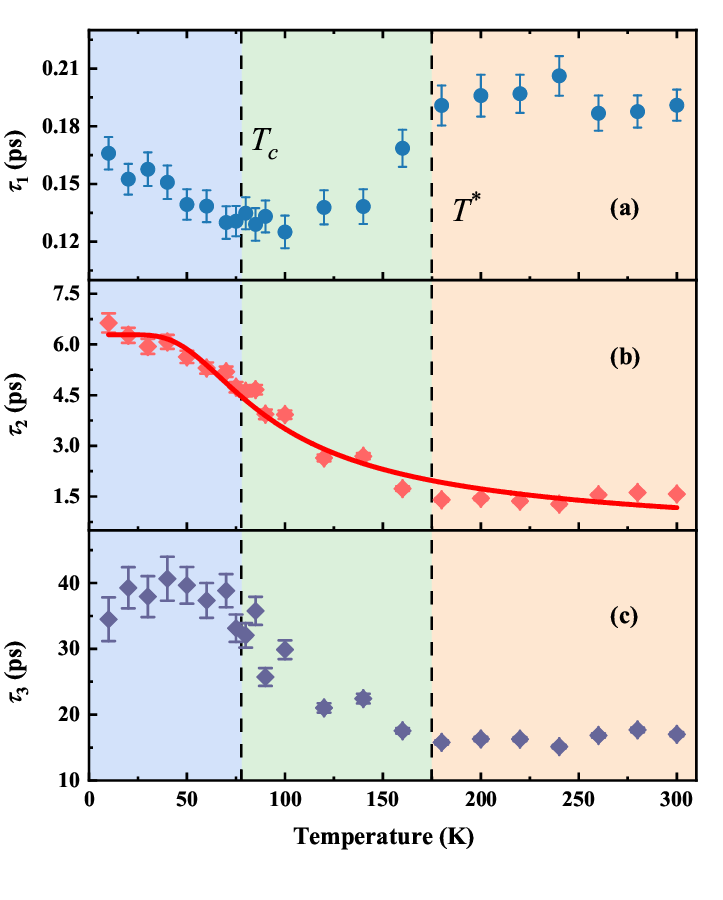}
	\subfigure{\label{4a}}
	\subfigure{\label{4b}}
	\subfigure{\label{4c}}
	\subfigure{\label{4d}}
	\caption{Temperature dependence of relaxation times (a) $\tau_{1}$, (b) $\tau_{2}$ and (c) $\tau_{3}$. The solid curve is the fitting result using Eq. \eqref{E2}.}
	\label{4}
\end{figure}

The temperature dependence of the extracted relaxation times is depicted in Fig. \ref{4}. All the three decay times displays very weak temperature dependence above $T^* \sim$ 175 K. $\tau_{2}$ and $\tau_{3}$ increase towards lower temperature, while $\tau_{1}$ first decreases with decreasing temperature, and then remains constant before it starts to increase below $T_c\sim$78 K. Even though the ferrimagnetic order in MST is established only at $T_c$, the spin fluctuation or short-range magnetic correlation was demonstrated to survive up to temperatures close to 2$T_c$ \cite{may2017magnetic,Spin160K}. Therefore, the gradual decrease of $\tau_{1}$ from 200 fs above $T^*$ to 120 fs around 100 K indicates that the onset of spin fluctuations provides an additional relaxation channel and thus decreases $\tau_{1}$ \cite{mijin2022spin}. With decreasing temperature below $T_c$, the formation of long-rang magnetic order will speed up the relaxation process via enhanced spin-phonon interaction \cite{2020Anomalous,cheng2023MBT}, resulting in a further decrease of  $\tau_{1}$. However, this is clearly the opposite of our observation, as shown in Fig. \ref{4a}. This discrepancy may be because the electron-electron thermalization slows down in the magnetic state and becomes comparable to or longer than the timescale of \eph thermalization at low temperature \cite{Demsar_2006,2020Anomalous}. Therefore, the fast decay time is governed by the electron thermalization times, which increase linearly with decreasing temperature.

Following \eph thermalization processes, hot carriers can still accumulate on the band edge and finally return to the unexcited equilibrium state though recombination of hot carriers. Radiative recombination usually occurs in direct bandgap semiconductors with a time constant in the nanosecond range, which is much longer than $\tau_{2}$ and $\tau_{3}$. In indirect bandgap semiconductors, \eh recombination is mediated by phonons or impurities in order to maintain momentum conservation. MST is a semiconductor with an indirect gap of 120 meV \cite{seo2021colossal,Qiong2023}. In particular, the phonon-assisted \eh recombination has been observed in several topological materials with indirect bandgaps, where the timescale and temperature dependence are very similar to $\tau_{2}$ as observed here \cite{dai2016ultrafast,LIU2021,cheng2022ultrafast}.It has been proposed that the temperature dependence of the phonon-assisted \eh recombination time can be quantitatively described by \cite{lopez1968electron}
\begin{equation}\label{E2}
	\frac{1}{\tau_2} = A\frac{x}{sinh^2x}+\frac{1}{\tau_d},
\end{equation}
where $A$ is a fitting parameter, $x=\hbar \omega/{2k_BT}$, $\omega$ is the frequency of the phonon mode that assists the \eph recombination, and  $\tau_d$ is the impurity scattering time which is independent of temperature but depends on the impurity density. The solid line in Fig. \ref{4b} represents the fitting result with the parameters $\omega/2\pi$ = 6.2 THz and $\tau_d$ = 6.3 ps. The good fitting quality supports the assignment of phonon-assisted \eh recombination to the $\tau_{2}$ relaxation process. According to the first-principles calculations, the 6.2 THz phonon is related to the vibrations of Mn and Si atoms \cite{Li_2023}. It is reasonable to expect that the decreased phonon population at low temperature slows down the phonon-assisted recombination process and leads to a longer relaxation time.

A similar temperature dependent behavior is observed in $\tau_{3}$ except the slight decrease at low temperature, as illustrated in Fig. \ref{4c}. However, it can not be attributed to the phonon-assisted \eh recombination since its fluence dependence is different from that of $\tau_{2}$. As shown in Figs. \ref{3e} and \ref{3f}, $\tau_{2}$ is independent of pump fluence, while $\tau_{3}$ decreases with increasing pump fluence. In addition, $\tau_{3}$ is roughly an order of magnitude larger than $\tau_{2}$. In MST, impurity bands close to the valence bands \cite{seo2021colossal}, which can act as non-radiative recombination centers exist. SRH recombination is a process in which electrons from the conduction band recombine with holes from the valence band through traps. The fluence-dependent lifetime related to the SRH mechanism can be described by \cite{SRH1952}
\begin{equation}\label{E3}
	\frac{1}{\tau_3} = \gamma_s\frac{1+cN_e}{1+aN_e},
\end{equation}
where $\gamma_s$ is the recombination rate at small photo-excitation, $c$ and $a$ are fitting parameters, and $N_e$ is the photo excited carrier density proportional to the pump fluence. As illustrated in Fig. \ref{3f}, the fitting results agree quite well with the extracted $\tau_3$. Since the SRH recombination time strongly depends on the relative position of trap states with respect to the Fermi level in thermal equilibrium \cite{SRH1952}, the rapid increase of $\tau_{3}$ from $\sim$15 ps above $T^*$ to the saturation value of $\sim$40 ps below $T_c$ is likely associated with the decrease of activation energy with increasing magnetic order, as evidenced by the transport experiments  \cite{may2017magnetic,seo2021colossal,Polaronic2021,Spin160K}.

\subsection{\label{sec:level2} Coherent phonons dynamics}

Next, we discuss the oscillatory components of \R that describe the phonon dynamics. The fast and slow oscillations can be attributed to coherent optical and acoustic phonons, respectively. Figure \ref{5a} shows the high-frequency oscillations after subtracting the electric response in the first 6 ps, and the temperature-dependent frequency domain data are plotted in Fig. \ref{5b}. Remarkably, a sharp peak around 3.7 THz and a broad hump with a center frequency near 4.5 THz were observed at room temperature. In the following, we mainly focus on the $f_1$ mode. As the temperature is lowered, the $f_1$ mode frequency increases while its decay rate decreases. Similar temperature-dependent behavior has been observed in several topological materials, where it was attributed to the decay of the coherent optical phonon into two counter-propagating acoustic phonons via phonon-phonon interaction \cite{T1984,Hot2020ZrTe5,cheng2014temperature}. The anharmonic phonon decay process can be described in the form \cite{qi2010ultrafast,zhu2021temperature}

\begin{align}
	{}f_{1}(T)= f_0+c_1\left[ 1+ 2n_B(\omega_0/2,T)\right] \label{E4}\\
	\gamma_{1}(T)= \gamma_0+c_2\left[ 1+ 2n_B(\omega_0/2,T)\right]  \label{E5}
\end{align}
where $f_{0}$ is the optical phonon frequency at 0 K, $\gamma_{0}$ is defect-induced damping term, and $c_1$ and $c_2$ are constants. $n_B(\omega,T)=\left(e^{\hbar\omega/k_BT}-1\right)^{-1}$ is the optical phonon occupation. The solid lines in Fig. \ref{5c} represent a simultaneous fitting of $f_{1}(T)$ and $\gamma_{1}(T)$, yielding parameters $f_0$ = 3.8 THz and $\gamma_{0}$ = 1.85 THz.  The good fitting quality indicates that the decay of the $f_1$ optical phonon mode is dominated by the three-phonon scattering process. However, this phonon mode does not match with the theoretically calculated and experimentally measured values in MST \cite{wang2022pressure,mijin2022spin,Li_2023}.

To further verify the origin of observed optical phonons, we performed cw Raman spectroscopy on another freshly cleaved MST crystal at room temperature using a Jobin Yvon LabRam HR800 spectrometer with 633 nm laser light. As shown in Fig. \ref{4d}, the Raman spectrum at low laser intensity shows three peaks at 2.85 THz (95.1 \cm), 3.2 THz (106.6 \cm), and 4.5 THz (150.6 \cm), which are assigned to the $E_g$ (the first peak) and \Ag phonons of bulk MST, consistent with previous Raman studies \cite{wang2022pressure, mijin2022spin}. However, an additional peak at $\sim$3.8 THz (126.8 \cm) appears, exhibiting a red shift with increased peak intensity as the excitation power increases. In addition, a shoulder near 4.3 THz (143.4 \cm) appears at a cw laser power of 300 $\muup$W. Such anomalous Raman modes were frequently observed in Te-based binary and ternary chalcogenides induced by the segregation of tellurium after high-fluence laser irradiation \cite{Ishioka_2006,Fukuda2022,Manjon2021}. Since MST has low lattice thermal conductivity \cite{Polaronic2021}, the heat generated by laser radiation cannot be dissipated efficiently, resulting in the segregation of Te even with cw excitation power of 84 $\muup$W in the Raman scattering measurement. Based on the comparison with Raman spectra, we ascribe the $f_1$ mode to the \Ag phonon of Te. In addition, the $E_g$ phonon of Te with a frequency of 4.3 THz was also observed in the transient reflectivity measurement of CdTe \cite{Ishioka_2006}. Therefore, we cannot entirely exclude the possibility that the $f_2$ mode originates from the presence of Te precipitates, although it agrees well with the out-of-plane \Ag optical phonon of MST. Note that the significant difference between the Raman and FFT spectra may be due to the different configuration of two measurements.

The acoustic phonon changes very little in the whole temperature range, as depicted in Fig. \ref{1b}. A frequency of the coherent acoustic phonon $f_{3}$ = 0.026 THz and a dephasing time of 209 ps are obtained from the fitting in Fig. \ref{2}. The acoustic phonon in transient reflectivity is related to the pump pulse generated strain wave propagating into the sample with sound velocity $v_s$ \cite{thomsen1986surface}. The oscillatory frequency is given by $f_{3}=2nv_scos\theta/\lambda_{probe}$, where $n$ is the refractive index, $\theta$ is the incidence angle of the probe beam, and $\lambda_{probe}$ is the probe wavelength. The penetration depth of the probe pulse can be described by  $\xi=v_s/\gamma_{3}=\lambda_{probe}/4\pi k$. In addition, the linear reflectivity of the probe pulse is separately measured to be a constant value of 0.4. Combining the above two expressions with the Fresnel formula, the refractive index $n$ and extinction coefficient $k$ can be self-consistently derived with the above formula \cite{kumar2012acoustic}. Correspondingly, the penetration depth $\xi$ and the sound speed $v_s$ are estimated to be 506 nm and 2.42 km/s at 300 K, respectively. The derived sound speed, which is comparable to the calculated value of the longitudinal acoustic phonons, is expected to lead to the low thermal conductivity in MST \cite{Li_2023}.

\begin{figure}[htbp]
	\includegraphics[width=2.2 in]{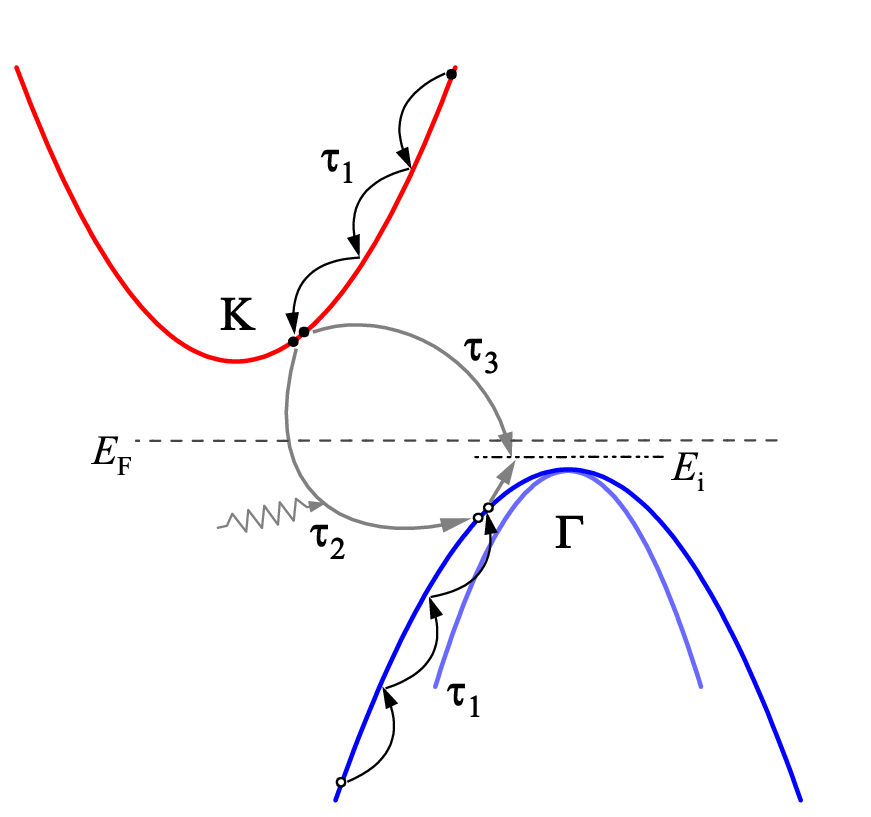}
	\caption{Schematic band structure of MST near the Fermi energy along the $\Gamma$-K direction. $\tau_{1}$, $\tau_{2}$ and $\tau_{3}$ represent \eph thermalization, phonon-assisted \eh recombination, and trap-assisted \eh recombination, respectively. $E_F$ and $E_i$ are the Fermi level and impurity level, respectively.}
	\label{6}
\end{figure}

\section{Conclusion}

Based on the above analysis, we summarize the quasiparticle relaxation dynamics in MST as schematically illustrated in Fig. \ref{6}. A small indirect band gap of $\sim$ 160 meV along the $\Gamma$-$K$ direction with nodal-line structure of the valence Te band below the Fermi level exists\cite{seo2021colossal,zhang2023electronic}. As the pump photon energy is higher than the bandgap of MST, the electrons are excited into the conduction band. The hot carriers transfer their excess energy to the lattice by emission of optical phonons on a timescale of $\tau_1$. However, under high-excitation conditions, the hot-phonon bottleneck effect slows the carrier cooling. After that, the electrons and holes recombine with the assistance of phonons ($\tau_2$) and the impurity states ($\tau_3$). At temperatures above $T^*$, all three relaxation times exhibit weak temperature dependence indicating the semiconducting nature. In the paramagnetic state below $T^*$, considerable in-plane spin fluctuation was demonstrated by neutron diffraction and Raman scattering measurements  \cite{may2017magnetic,Spin160K,mijin2022spin}. However, the long-range ferrimagnetic order is established only below $T_c$. Since the electronic structure of MST strongly depends on the spin orientation, the temperature evolution of the magnetic order influences the electronic structure, which in turn affects the carrier relaxation. 

In summary, we have utilized ultrafast optical spectroscopy to investigate the nonequilibrium carrier and coherent phonon dynamics in the ferrimagnetic semiconductor Mn$_3$Si$_2$Te$_6$. Quantitative analysis revealed that the \eph thermalization timescale is prolonged due to the hot phonon bottleneck effect. We identified two non-radiative recombination processes, i.e., phonon-assisted \eh recombination and trap-assisted SRH recombination. Significantly, temperature-dependent measurements showed that all three relaxation components show large variation around 175 and 78 K, indicating the complicated magnetic interaction in Mn$_3$Si$_2$Te$_6$. In addition, we observed two distinct optical phonons and one pronounced acoustic phonon. The temperature dependence of the frequency and damping rate of the $f_1$ mode can be well described by an anharmonic phonon model. In combination with the Raman spectra, we ascribe this phonon mode to the $A_{1g}$ Raman mode of segregated Te. Finally, we estimated a penetration depth of 506 nm and a sound speed of 2.42 km/s in MST for 800 nm laser light by applying the strain pulse propagation model to the coherent acoustic phonon. Our findings develop the understanding of the nonequilibrium properties of the ferrimagnetic semiconductor MST and also shed light on its potential applications in ultrafast devices.

\begin{acknowledgments}
This work was supported by the National Natural Science Foundation of China (Grants No. 11974414 and No. 12204520) and the Synergetic Extreme Condition User Facility (SECUF). Z.L.L. is grateful for the support from the Youth Innovation Promotion Association, CAS (Grant No. 2021008).
\end{acknowledgments}

\bibliographystyle{apsrev4-2}
\bibliography{MST_ref.bib}

\end{document}